\documentclass[twocolumn,english,prl]{revtex4-2}
\usepackage[latin9]{inputenc}
\usepackage{color}
\usepackage{amsmath}
\usepackage{amssymb}
\usepackage{graphicx}
\usepackage{esint}

\makeatletter
\usepackage{xcolor}

\makeatother

\usepackage{babel}
\begin{document}
\title{Surface Luttinger arcs in Weyl semimetals}
\author{Osakpolor Eki Obakpolor, Pavan Hosur}
\affiliation{University of Houston, Houston, TX 77204}
\begin{abstract}
The surface of a Weyl semimetal famously hosts an exotic topological
metal that contains open Fermi arcs rather than closed Fermi surfaces.
In this work, we show that the surface is also endowed with a feature
normally associated with strongly interacting systems, namely, Luttinger
arcs, defined as zeros of the electron Green's function. The Luttinger
arcs connect surface projections of Weyl nodes of opposite chirality
and form closed loops with the Fermi arcs when the Weyl nodes are
undoped. Upon doping, the ends of the Fermi and Luttinger arcs separate
and the intervening regions get filled by surface projections of bulk
Fermi surfaces. Remarkably, unlike Luttinger contours in strongly
interacting systems, the precise shape of the Luttinger arcs can be
determined experimentally by removing a surface layer. We use this
principle to sketch the Luttinger arcs for Co and Sn terminations
in Co$_{3}$Sn$_{2}$S$_{2}$. The area enclosed by the Fermi and
Luttinger arcs approximately equals the surface particle density in
weakly coupled systems while the correction is governed by the interlayer
couplings and the perimeter of the Fermi-Luttinger loop.
\end{abstract}
\maketitle
The past decade has seen tremendous advancements in the field of Weyl
semimetals (WSMs) -- three-dimensional (3D) materials defined by
the presence of non-degenerate Bloch electron bands that intersect
at arbitrary points in the Brillouin zone \citep{VafekDiracReview,Burkov2018,Burkov:2016aa,YanFelserReview,ArmitageWeylDiracReview,Shen2017,Belopolski:2016wu,Guo2018,Chang2016,Gyenis_2016,Huang:2015vn,Inoue1184,Lv:2015aa,Sun2015a,Xu2015,Xu2016,Xu613,Yang:2015aa,Zheng2016}.
Near the intersection points, or Weyl nodes, the Bloch dispersion
resembles the Weyl dispersion that is well-known in high-energy physics;
hence, the name WSM. The Weyl nodes carry a chirality or handedness
of $\pm1$, emit or absorb unit Berry flux depending on their chirality,
and are even in number such that the total chirality vanishes. Moreover,
the nodes are topological, in the sense that they exhibit profound
topological quantum anomalies, can only be destroyed by annihilating
in pairs of opposite chirality and spawn a myriad of topological transport
phenomena \citep{Hosur2013a,Wang_2018,Hu:2019aa,ZyuninBurkovWeylTheta,ChenAxionResponse,VazifehEMResponse,Burkov_2015,Hosur2012,Juan:2017aa,Wang2017,Nagaosa:2020aa,NielsenABJ,IsachenkovCME,SadofyevChiralHydroNotes,Loganayagam2012,GoswamiFieldTheory,Wang2013,BasarTriangleAnomaly,LandsteinerAnomaly}.

WSMs famously host metallic surface states known as Fermi arcs (FAs)
that connect projections of Weyl nodes of opposite chirality onto
the surface Brillouin zone \citep{Benito-Matias2019,Chang2016,Gyenis_2016,Deng2017,Deng:2016aa,Guo2018,HaldaneFermiArc,Hosur2012a,Huang2016,Huang:2015vn,Iaia:2018aa,Inoue1184,Kwon:2020aa,Lau2017,Sakano2017,Sun2015a,Lv:2015aa,Xu2015,Xu2015a,Xu2015b,Xu2016,Xu613,XuLiu2018,Yuan2018QPI,Zhang:2017ac,Yuaneaaw9485,Moll:2016aa,Potter2014,Zhang2016}.
They are mandated by the bulk-boundary correspondence in topological
matter analogous to the Dirac cone surface states in topological insulators
\citep{HasanKaneReview,QiZhangRMP,MongShivamoggiEdge}. However, the
latter are exponentially localized near the surface whereas the bulk
penetration depth of the FA wavefunction depends strongly on its surface
momentum $\boldsymbol{k}$ and diverges at its end-points. The metallic
nature of FAs can manifest in several ways such as quantum oscillations
due to peculiar cyclotron orbits in mixed real and momentum space
\citep{Borchmann2017,Moll:2016aa,Potter2014,Zhang2016,Zhang:2017ac,Zhang2019trefoil,Nishihaya:2021um,Zhang:2019ab},
unusual collective modes due to mixing between FAs and bulk states
\citep{Bonacic2018,Bugaiko2020,Gorbar2019,Hofmann2016,Tamaya_2019,Song2017,Andolina2018,Adinehvand2019,Ghosh2020}
and susceptibility of the surface to proximity-induced superconductivity
\citep{Faraei2020,huang2019,Khanna2014}. However, the absence of
a closed Fermi surface in the FA metal renders the Luttinger's theorem
-- a fundamental equality between the particle density and the volume
within the Fermi surface in Fermi liquids \citep{Blagoev1997,Luttinger1960,Oshikawa2000,Seki2017,Voit_1995,Yamanaka1997,Heath_2020}
-- inapplicable.

In this work, we show that surface of a WSM also hosts Luttinger arcs
(LAs), defined as momentum space regions where the electron Green's
function vanishes. LAs and Luttinger surfaces are known to occur in
strongly interacting systems due to vanishing quasiparticle weight
or diverging self-energy \citep{Dzyaloshinskii2003,Seki2017,Stanescu2007,Yamanaka1997,Heath_2020}.
Heuristically, LAs in WSMs can be viewed as a manifestation of strong
self-interactions among surface electrons mediated by bulk states.
We emphasize, however, that the system is strictly non-interacting.
Interestingly, when all the Weyl nodes are undoped, the FAs and LAs
form closed loops. Moreover, LAs transform into FAs when a surface
layer is removed, which enables their detection. Using this idea,
we determine the LAs on the (001) surface of the ferromagnetic WSM
Co$_{3}$Sn$_{2}$S$_{2}$ \citep{Okamura:2020tq,Liu2019Co3Sn2S2,Morali2019,XuLiu2018,Guin:2019tx}
based on recent scanning tunneling data \citep{Morali2019}.

\emph{Surface Green's function:} Let $\tilde{H}_{\boldsymbol{k}}$
denote the Bloch Hamiltonian of an $L$-layered system that has $D_{z}$
degrees of freedom in the $z^{th}$ layer. The layers are unrelated
in general, but repeat periodically in lattice models. As usual, the
full single-particle Green's function is $\tilde{G}_{\boldsymbol{k}}(\omega)=\left(\omega-\tilde{H}_{\boldsymbol{k}}\right)^{-1}$
and has $\tilde{N}=\sum_{z=1}^{L}D_{z}$ poles corresponding to the
eigenvalues $\tilde{\varepsilon}_{\boldsymbol{k},j}$, $j=1\dots\tilde{N}$
of $\tilde{H}_{\boldsymbol{k}}$. The spectrum may have degeneracies
and generically consists of three types of states: evanescent waves
pinned to the top and the bottom surfaces and planes waves in the
bulk. 

Now, let us add a layer at $z=0$ that we will refer to as the ``surface''.
The full and effective surface Green's function, respectively, are
given by
\begin{align}
G_{\boldsymbol{k}}(\omega) & =\left(\begin{array}{cc}
\omega-H_{\boldsymbol{k}}^{S} & -h_{\boldsymbol{k}}\\
-h_{\boldsymbol{k}}^{\dagger} & \omega-\tilde{H}_{\boldsymbol{k}}
\end{array}\right)^{-1}\\
g_{\boldsymbol{k}}(\omega) & =\left(\omega-H_{\boldsymbol{k}}^{S}-h_{\boldsymbol{k}}\tilde{G}_{\boldsymbol{k}}(\omega)h_{\boldsymbol{k}}^{\dagger}\right)^{-1}
\end{align}
where $H_{\boldsymbol{k}}^{S}$ is the Bloch Hamiltonian of the surface
layer and $h_{\boldsymbol{k}},h_{\boldsymbol{k}}^{\dagger}$ capture
coupling between the bulk and the surface and $g_{\boldsymbol{k}}(\omega)$
is a matrix of dimension $D_{0}$. A standard identity for the determinant
of a block matrix gives 
\begin{align}
\det g_{\boldsymbol{k}}(\omega) & =\frac{\det G_{\boldsymbol{k}}(\omega)}{\det\tilde{G}_{\boldsymbol{k}}(\omega)}=\frac{\prod_{i=1}^{\tilde{N}}\left(\omega-\tilde{\varepsilon}_{\boldsymbol{k},i}\right)}{\prod_{j=1}^{N}\left(\omega-\varepsilon_{\boldsymbol{k},j}\right)}\label{eq:det-g}
\end{align}
Moreover, if interlayer hopping is sufficiently local, $\tilde{H}_{\boldsymbol{k}}$
and $H_{\boldsymbol{k}}$ must have the same spectrum of evanescent
waves on the $z=L$ surface upto exponentially small corrections in
$L$. Thus, factors from such states cancel out in Eq. (\ref{eq:det-g}),
leaving $\det g_{\boldsymbol{k}}(\omega)$ to only depend on the bulk
and top-surface spectra of $H_{\boldsymbol{k}}$ and $\tilde{H}_{\boldsymbol{k}}$.
Similarly to single-particle Green's functions in interacting systems
\cite{Seki2017}, $\det g_{\boldsymbol{k}}(\omega)$ is a ratio of
zeros to poles, vanishes as $|\omega|\to\infty$ and is analytic away
from the $\text{Re}(\omega)$ axis.

\begin{figure}
\includegraphics[width=0.8\columnwidth]{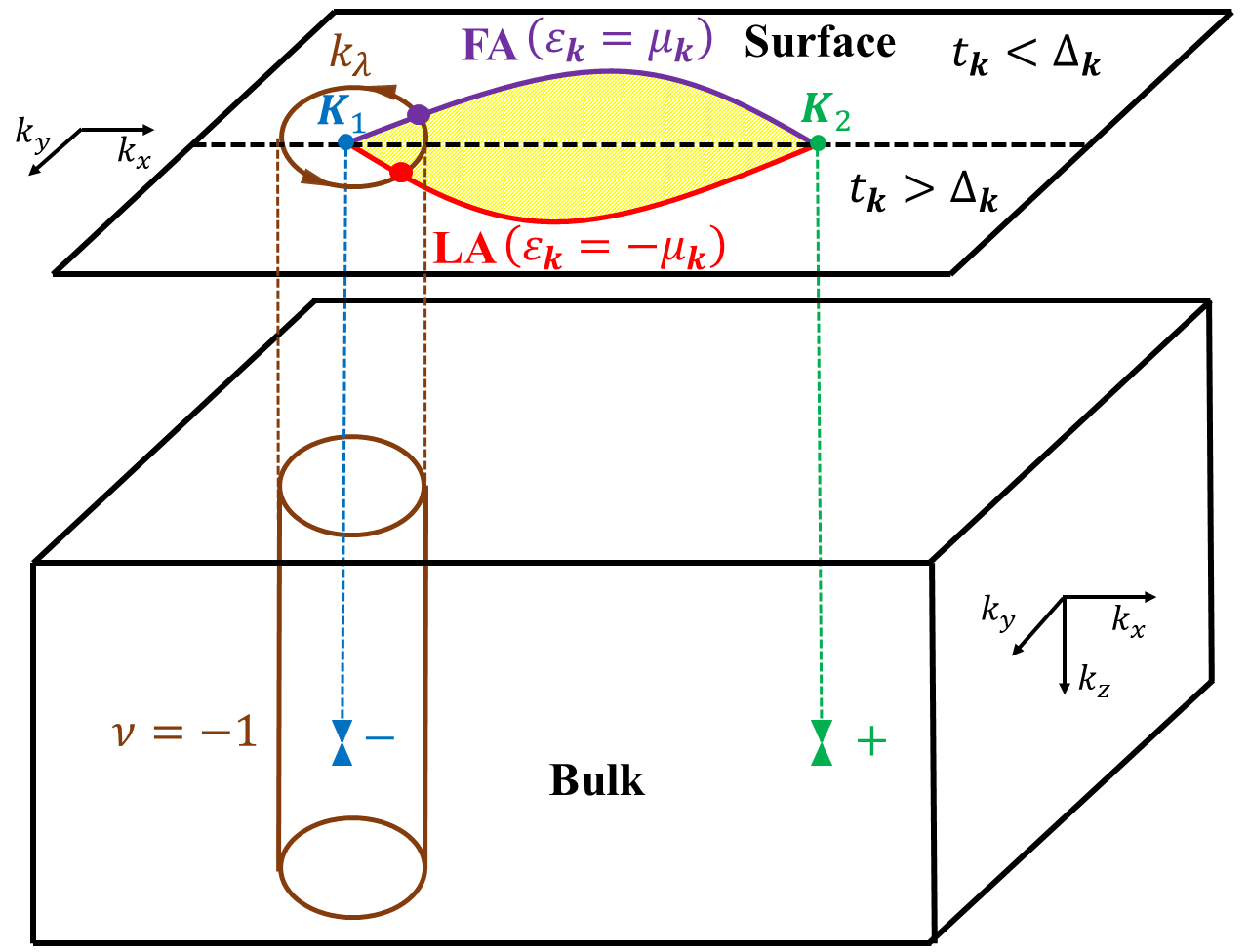}

\caption{Blue (green) cone and dot denotes the left-(right-)handed bulk Weyl
node and its surface projection, respectively. While traversing a
surface $\boldsymbol{k}$-space loop (brown curve parametrized by
$k_{\lambda}$), the phase acquired by $g_{k_{\lambda}}(i0^{+})$
on crossing the FA (purple curve) must be canceled by crossing a LA
(red curve). The FA and LA form a closed and enclose a well-defined
area (yellow region). The bulk extension of the $k_{\lambda}$-curve
defines a 2D insulator on the $k_{\lambda}$-$k_{z}$ manifold (brown
cylinder) with Chern number equal to the net chirality of the enclosed
Weyl nodes. $\mathcal{E}_{\boldsymbol{k}}$, $\mu_{\boldsymbol{k}}$,
$t_{\boldsymbol{k}}$ and $\Delta_{\boldsymbol{k}}$ are parameters
in the explicit model described later. \label{fig:k-loops-mu-0}}
\end{figure}

\emph{Luttinger arcs:- }We now prove our main results, namely, the
existence, connectivity and detection of surface LAs by merely requiring
$\det g_{\boldsymbol{k}}(\omega)$ to be single-valued. In particular,
we prove that LAs (i) necessarily exist on the surface; (ii) connect
Weyl node projections of opposite chirality, thus forming closed loops
with the FAs; and (iii) can be precisely determined by peeling off
suitable layers and measuring the new FAs. First, we derive the well-known
existence of the FAs within a setup that facilitates the proof for
LAs. 

Consider a momentum space loop in the surface Brillouin zone, parametrized
by $k_{\lambda}$, that does not pass through the surface projection
of any Weyl node (Fig. \ref{fig:k-loops-mu-0}). In the bulk, the
surface defined by $k_{\lambda}$ and $k_{z}$ is a closed 2D manifold.
If the Weyl nodes are undoped, the manifold has a gapped spectrum
and can be viewed as a 2D Chern insulator with a Chern number $\nu$
equal to the net chirality of the enclosed Weyl nodes. The edge of
this Chern insulator is the original $k_{\lambda}$-loop on the WSM
surface, so the loop must host a net of $\nu$ gapless, chiral modes,
where each clockwise (counter-clockwise) mode contributes $+1$ ($-1$)
to $\nu$. In other words, each right-(left-)handed Weyl node produces
a FA state with a velocity component along the clockwise (counter-clockwise)
direction when viewed from above. In terms of Green's functions, each
FA state corresponds to a vanishing non-degenerate eigenenergy $\varepsilon_{k_{\lambda},j}$
for a certain $j$ and contributes a simple pole to $g_{\boldsymbol{k}}(\omega)$
at $\omega=0$. Moreover, the \emph{net }number of poles equals $\nu$,
where FAs with a velocity component anti-parallel (parallel) to the
loop direction contribute $+1$ ($-1$). The set of all possible loops
on the WSM surface then traces out the FAs.

\begin{figure}
\begin{centering}
\includegraphics[width=0.4\columnwidth]{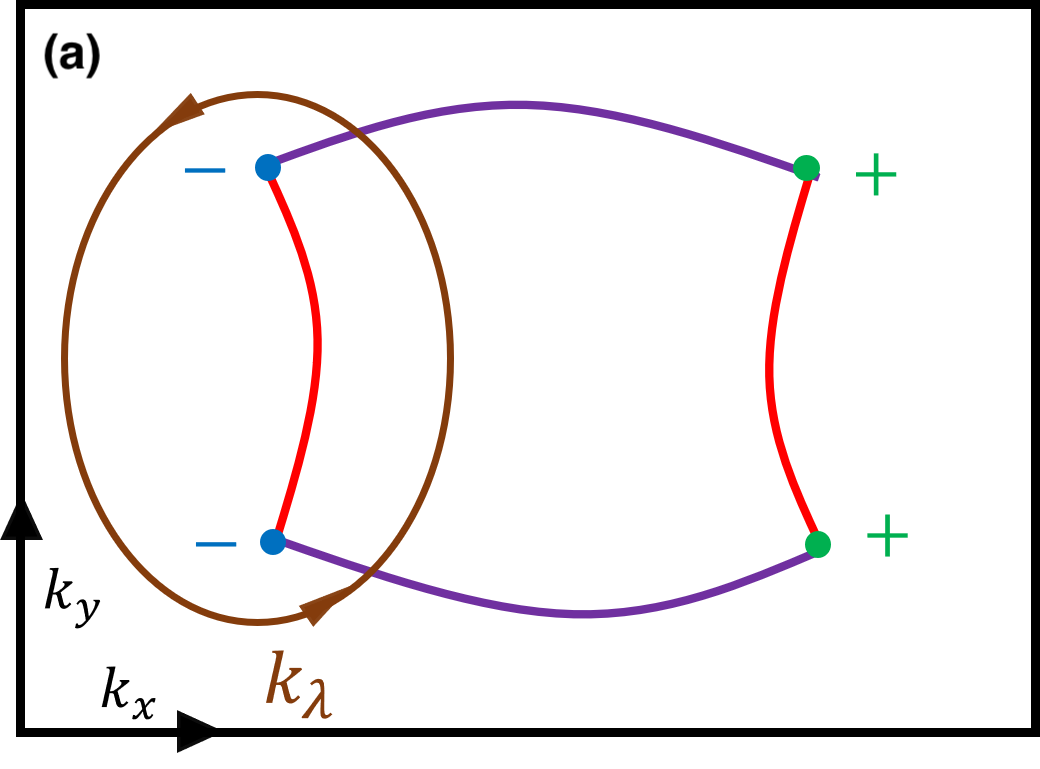}\includegraphics[width=0.4\columnwidth]{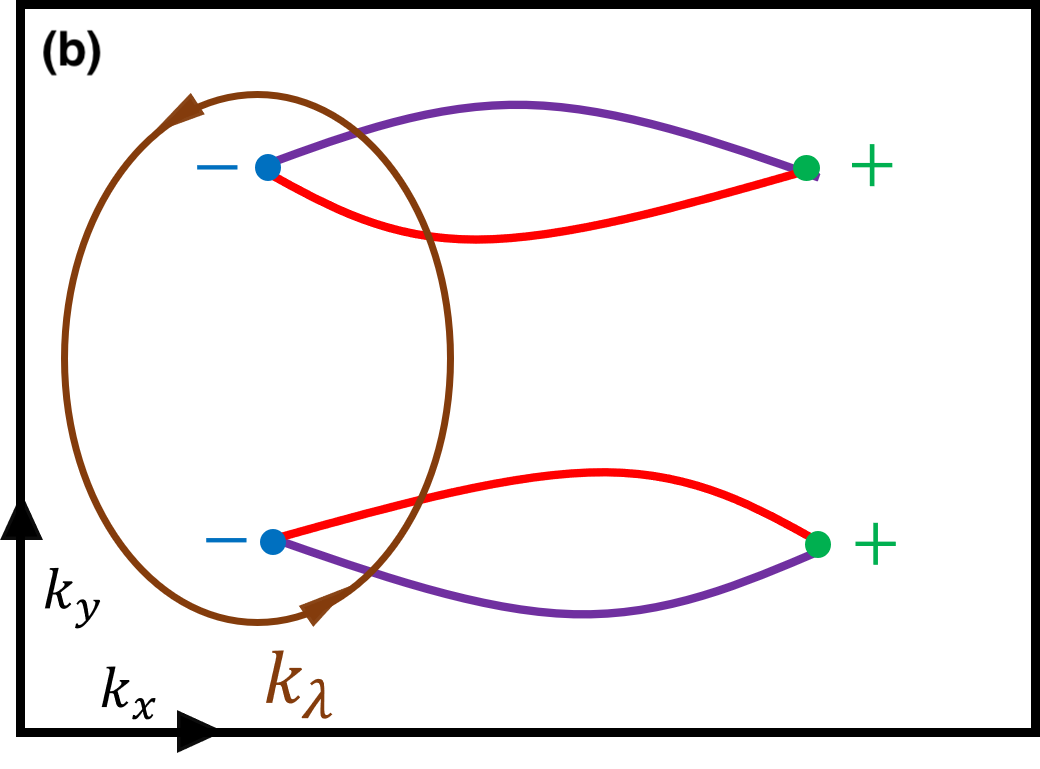}
\par\end{centering}
\caption{Invalid (a) and valid (b) LA-configurations, where LAs connects surface
projections of Weyl nodes of the same and opposite chiralities, respectively.
\textcolor{red}{\label{fig:LA-configurations}}}
\end{figure}

Now, consider the retarded surface Green's function, $g_{\boldsymbol{k}}(\omega+i0^{+})$.
A necessary condition for it to be single-valued is that the change
in $\ln\det g_{\boldsymbol{k}}(\omega+i0^{+})$ vanishes around an
arbitrary $k_{\lambda}$-loop. Focusing on $\omega=0$ and using Eq.
(\ref{eq:det-g}),
\begin{align}
0 & =\oint d\boldsymbol{k}_{\lambda}\cdot\boldsymbol{\nabla}\ln\det\left[g_{\boldsymbol{k}}(i0^{+})\right]\label{eq:loop integral}\\
 & =\pi\left(\sum_{m}\text{sgn}\left[\tilde{v}_{k_{m}^{LA}}\right]-\sum_{n}\text{sgn}\left[v_{k_{n}^{FA}}\right]\right)\nonumber 
\end{align}
where $k_{m}^{LA}$ $\left(k_{n}^{FA}\right)$ are points on the $k_{\lambda}$-loop
where $\tilde{\varepsilon}_{\boldsymbol{k},i}$ $\left(\varepsilon_{\boldsymbol{k},j}\right)$
vanishes for some $i$ ($j$), while $\tilde{v}_{k}$ $\left(v_{k}\right)$
is the projection of $\boldsymbol{\nabla}\tilde{\varepsilon}_{\boldsymbol{k},j}$
$\left(\boldsymbol{\nabla}\varepsilon_{\boldsymbol{k},j}\right)$
onto the loop direction. According to Eq. (\ref{eq:det-g}), $\det g_{\boldsymbol{k}}(0)$
has zeros (poles) at these points. While the poles correspond to FAs
and ensure $\sum_{n}\text{sgn}\left[v_{k_{n}^{FA}}\right]=-\nu$ as
argued above, Eq. (\ref{eq:loop integral}) implies that $\sum_{m}\text{sgn}\left[\tilde{v}_{k_{m}^{LA}}\right]=-\nu$
as well. In other words, there exist net $\nu$ zeros along the $k_{\lambda}$-loop.
The set of all $\boldsymbol{k}$-space loops yield strings of zeros,
that are defined as the LAs. Intuitively, the phase acquired by $\det\left[g_{\boldsymbol{k}}(i0^{+})\right]$
upon crossing a FA must be canceled either by a FA of the opposite
chirality or a LA of the same chirality to ensure $\det g_{\boldsymbol{k}}(i0^{+})$
is single-valued. This implies that the LAs too connect surface projections
of Weyl nodes of opposite chiralities and form closed loops with the
FAs when the Weyl nodes are undoped. To illustrate this point, we
show an invalid and a valid configuration in Fig. \ref{fig:LA-configurations}.

\emph{Detection by peeling:-} While Luttinger surfaces and arcs are
common in strongly interacting systems, their shape is difficult to
determine. There, a broad spectral function peak and low peak height
reflect diverging self-energy and vanishing quasiparticle weight,
respectively, but the precise point where the Green's function vanishes
is inaccessible \citep{Stanescu2007}. In contrast, we show that surface
LAs in WSMs can be determined by simply peeling off the top layer
and measuring the FAs on the new surface.

Suppose the $L$-layered system carries a FA that disappears upon
adding another layer. This can only happen if the unit cell consists
of more than one layer; otherwise the $L$- and $(L+1)$-layered systems
would have the same FA spectrum as $L\to\infty$. The disappearance
of FAs on adding a layer means there exists a string of $\boldsymbol{k}$
points such that $\tilde{\varepsilon}_{\boldsymbol{k},i}=0$ for some
$i$ but $\varepsilon_{\boldsymbol{k},j}\neq0\forall j$. According
to Eq. (\ref{eq:det-g}), $\det g_{\boldsymbol{k}}(0)=0$ along this
curve, thus yielding a LA on the surface. Equivalently, the LA transforms
into a FA when the surface layer is peeled off.

\begin{figure}
\includegraphics[width=0.8\columnwidth]{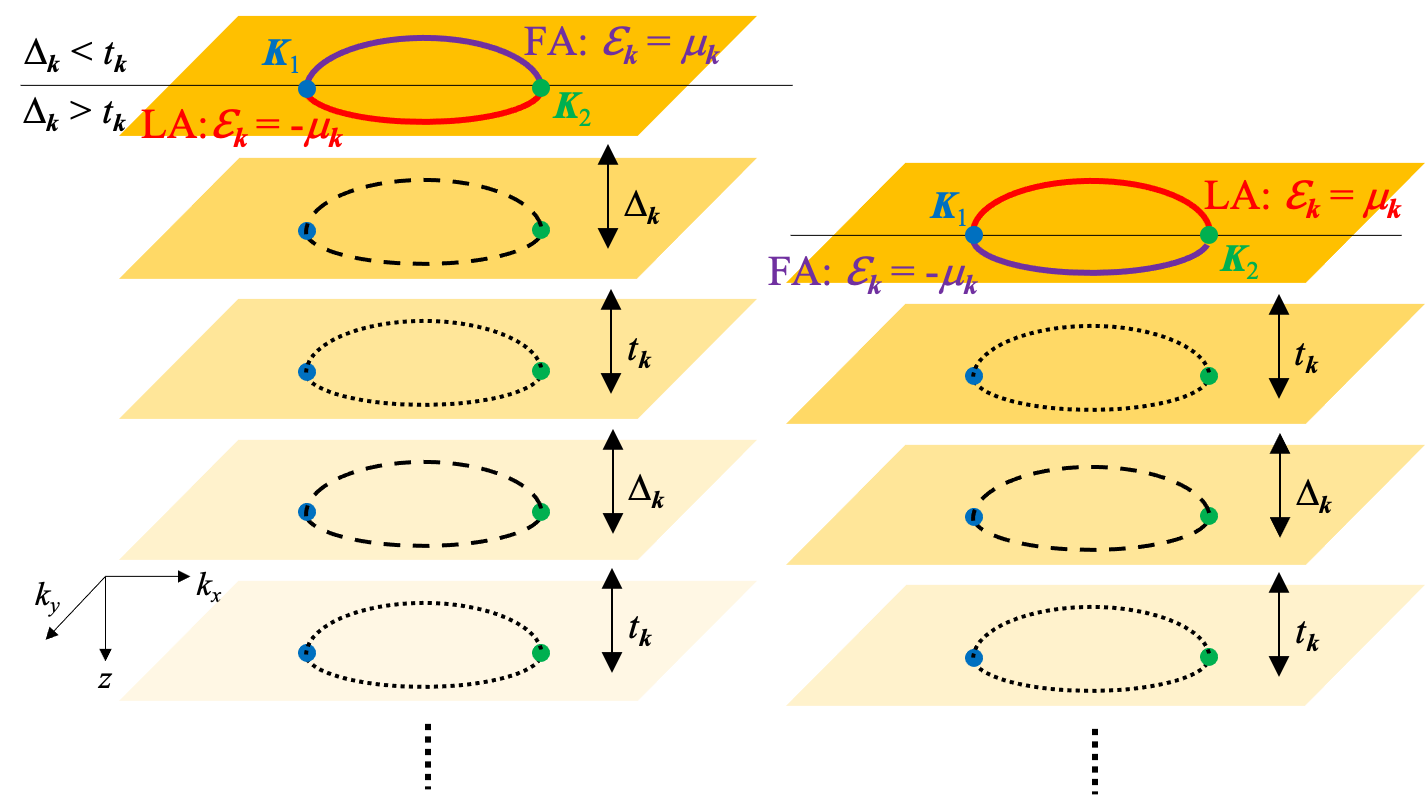}

\caption{In the bilayered WSM model, removing a layer interchanges the FA and
the LA. Dashed (dotted) black curves denote electron (hole) Fermi
surfaces that are the building blocks of the WSM (adapted from \citep{Hosur2012a};
see text for details).\label{fig:bilayer model}}
\end{figure}

This principle readily reveals the locations of LAs on the surface
of the ferromagnetic WSM Co$_{3}$Sn$_{2}$S$_{2}$. The crystal structure
of Co$_{3}$Sn$_{2}$S$_{2}$ consists of Co kagome layers with a
Sn at hexagon center separated by triangular layers of Sn and S \citep{Okamura:2020tq,Liu2019Co3Sn2S2,Morali2019,XuLiu2018,Guin:2019tx}.
Only three spinful Co $d$-orbitals from the three kagome sites and
a spinful $p$-orbital from the Sn atoms between the kagome layers
are near the Fermi level; bands from all other atoms and orbitals
are far away in energy. As a result, the material is effectively a
stack of bilayers consisting of Co-kagome and Sn-triangular layers.
Recent tunneling measurements on the (001) surface discovered well-isolated
FAs for Co and Sn terminations but with different connectivities \citep{Morali2019}.
We predict that the LAs on the Co (Sn) termination are simply the
FAs on the Sn (Co) termination (Fig. \ref{fig:bilayer model}). This
result is immune to the detailed orbital content of wavefunctions
and simply arises from the fact that LAs appear whenever FAs are annihilated
by adding a layer.

In contrast, the peeling protocol cannot reveal LAs in the antiferromagnetic
WSMs Mn$_{3}$Sn and Mn$_{3}$Ge \cite{Yang_2017,Matsuda:2020aa,Tomoya2018,Chen:2021uv,WangMn3Ge2021,Kubler2017}.
These materials have a layered structure where each layer consists
of a kagome lattice of Mn atoms with a Sn/Ge at the center of each
kagome hexagon. Importantly, the layers are idential up to inplane
translations in real space, so termination at any layer results in
the same surface Green's function in $\boldsymbol{k}$-space. 

\begin{figure}
\includegraphics[width=0.8\columnwidth]{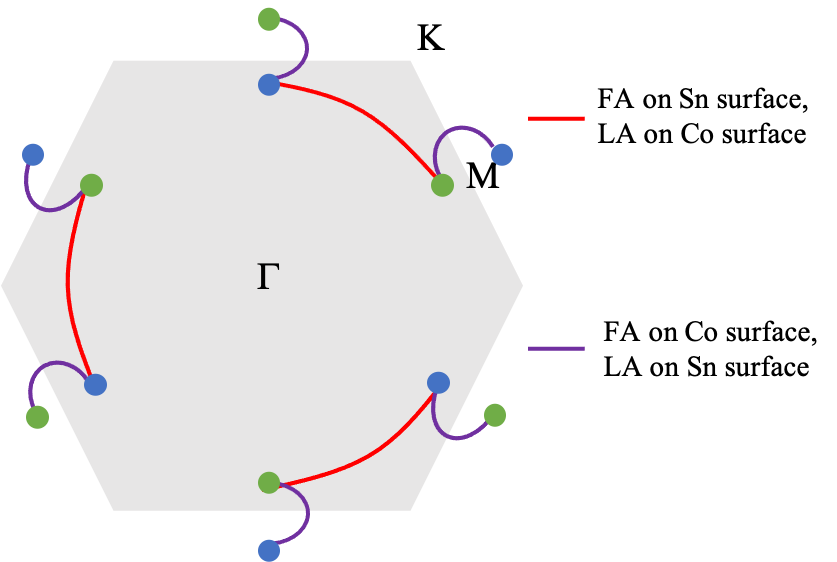}

\caption{Schematic of FAs and LAs on the surface of Co$_{3}$Sn$_{2}$S$_{2}$
for Co and Sn terminations. Gray region denotes the first Brillouin
zone while blue and green dots are surface projections of Weyl nodes
of opposite chirality.\label{fig:CoSnS-surface}}
\end{figure}

\emph{Effect of doping:-} So far, we have assumed every Weyl node
to be at charge neutrality. Real WSMs typically contain Fermi pockets
around Weyl nodes, which motivates an examination of the LAs under
doping.

We first need to analyze the effect of the bulk states on $g_{\boldsymbol{k}}(\omega)$
more closely. When $L\to\infty$, the bulk spectrum at each 2D momentum
$\boldsymbol{k}$ is continuous and indexed by $k_{z}$ while both
$\tilde{G}_{\boldsymbol{k}}(\omega)$ and $G_{\boldsymbol{k}}(\omega)$
contain a continuum of poles on the $\text{Re}(\omega)$ axis across
the energy range spanned by the bulk bands at $\boldsymbol{k}$. Naïvely,
one might expect the poles to cancel in Eq. (\ref{eq:det-g}) and
leave $g_{\boldsymbol{k}}(\omega)$ without any additional features.
However, for any finite $L$, the poles of $\tilde{G}_{\boldsymbol{k}}(\omega)$
and $G_{\boldsymbol{k}}(\omega)$ are generally distinct and typically
separated by $O(1/L)$ due to $k_{z}$-quantization. Crucially, $G_{\boldsymbol{k}}(\omega)$
has precisely $N-\tilde{N}=D_{0}$ more poles than $\tilde{G}_{\boldsymbol{k}}(\omega)$
does, $D_{0}$ being the number of degrees of freedom in the $z=0$
layer. Thus, even in the limit $L\to\infty$, the contour integral
$\oint d\omega\det g_{\boldsymbol{k}}(\omega)$ around the $\text{Re}(\omega)$
axis is generically non-zero and finite, which suggests that $g_{\boldsymbol{k}}(\omega)$
develops branch cuts that span the bulk bands.

\begin{figure}
\includegraphics[width=0.8\columnwidth]{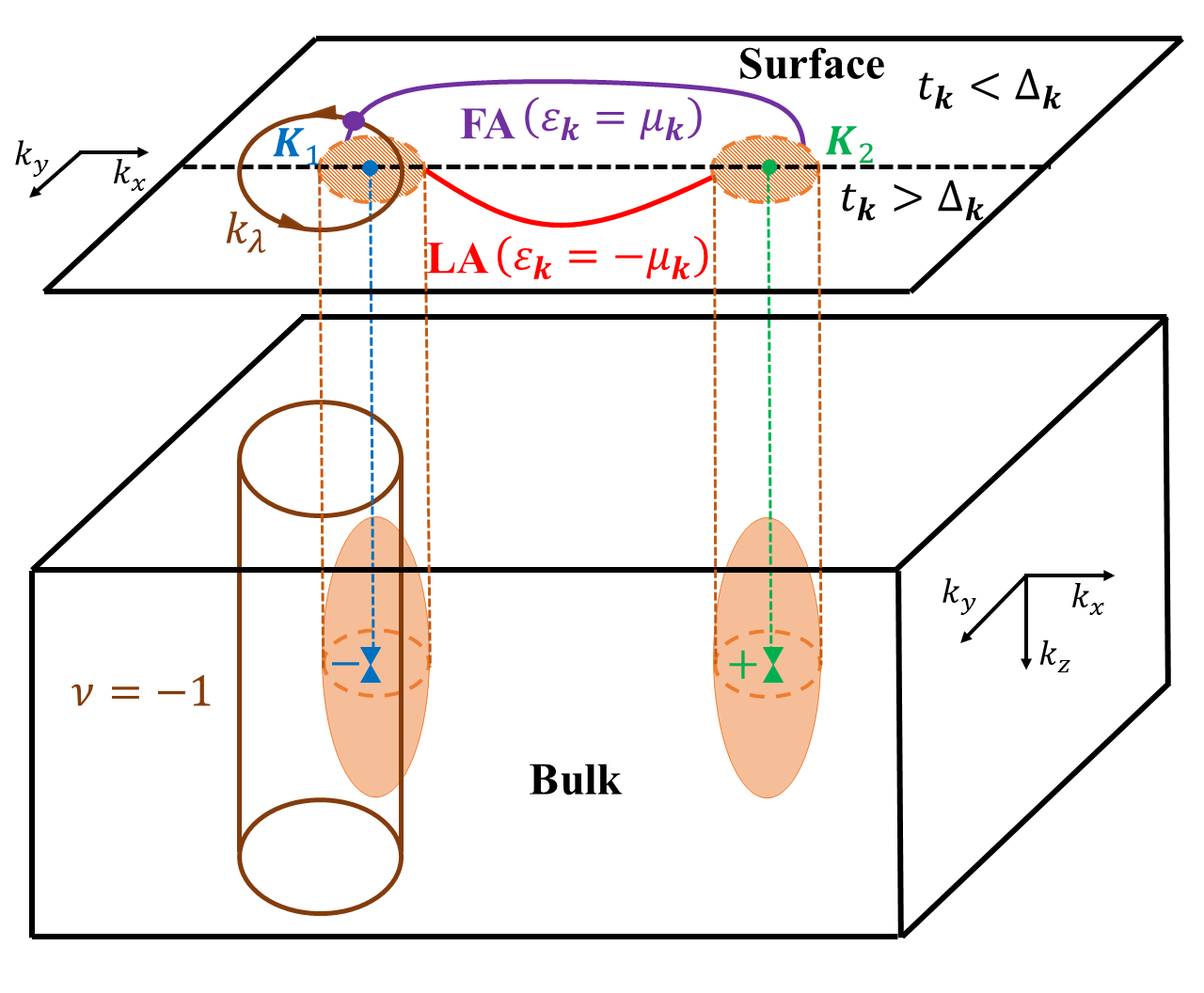}

\caption{Surface loops parametrized by $k_{\lambda}$ exist that intersect
the FA but avoid the LA. Along such a loop, the phase acquired by
$\det g_{\boldsymbol{k}}(i0^{+})$ on crossing the FA is canceled
by traversing a patch where $\text{Im}\det g_{\boldsymbol{k}}(i0^{+})$
is finite. These patches (orange ellipses) are the surface projections
of the bulk Fermi surfaces (orange ellipsoids). \label{fig:k-loops-mu-nonzero}}
\end{figure}

On the surface, doping creates patches of gapless states corresponding
to surface projections of the bulk Fermi surfaces (Fig. \ref{fig:k-loops-mu-nonzero}).
Consider a $k_{\lambda}$-loop that intersects a FA and then crosses
such a patch. When $\boldsymbol{k}$ is inside the patch, $\omega=0$
lies within a branch cut of $g_{\boldsymbol{k}}(\omega)$, so $\text{Im}\det g_{\boldsymbol{k}}(i0^{+})$
is finite. Consequently, the sign change in $\det g_{\boldsymbol{k}}(0)$
upon crossing the FA gets gradually undone while traversing the patch
without $\det g_{\boldsymbol{k}}(0)$ vanishing anywhere along the
loop. This causes the LA and the FA to separate while the patch fills
the intervening region.

\emph{Explicit model:-} We now demonstrate our general results using
an explicit model \citep{Hosur2012a}. The model consists of a stack
of spinless, alternating 2D electron and hole metals with Fermi surfaces
given by phenomenological curves $\mathcal{E}_{\boldsymbol{k}}=\pm\mu_{\boldsymbol{k}}$.
Interlayer couplings $-t_{\boldsymbol{k}}<0$ and $\Delta_{\boldsymbol{k}}>0$
cause each Fermi surface to hybridize preferentially with a layer
above for certain $\boldsymbol{k}$ and a layer below for other $\boldsymbol{k}$.
The Bloch Hamiltonian operator is $\hat{H}_{\boldsymbol{k}}=\sum_{z=0}^{L-1}\psi_{z,\boldsymbol{k}}^{\dagger}\left[-\mu_{\boldsymbol{k}}+(-1)^{z-1}\mathcal{E}_{\boldsymbol{k}}\right]\psi_{z,\boldsymbol{k}}+\left(\psi_{z,\boldsymbol{k}}^{\dagger}h_{zz+1,\boldsymbol{k}}\psi_{z+1,\boldsymbol{k}}+\textrm{h.c.}\right)$,
where $\psi_{z,\boldsymbol{k}}^{\dagger}$ creates an electron at
2D momentum $\boldsymbol{k}$ in the $z^{th}$ layer and $h_{zz+1,\boldsymbol{k}}$
equals $\Delta_{\boldsymbol{k}}$ ($-t_{\boldsymbol{k}}$) for even
(odd) $z$. The bulk Bloch Hamiltonian and dispersion are

\begin{align}
\mathcal{H}_{\boldsymbol{k}}(k_{z}) & =\left(\begin{array}{cc}
-\mu_{\boldsymbol{k}}+\mathcal{E}_{\boldsymbol{k}} & \Delta_{\boldsymbol{k}}-t_{\boldsymbol{k}}e^{2ik_{z}c}\\
\Delta_{\boldsymbol{k}}-t_{\boldsymbol{k}}e^{-2ik_{z}c} & -\mu_{\boldsymbol{k}}-\mathcal{E}_{\boldsymbol{k}}
\end{array}\right)\\
\xi_{\boldsymbol{k}}^{\pm}(k_{z}) & =-\mu_{\boldsymbol{k}}\pm\sqrt{\mathcal{E}_{\boldsymbol{k}}^{2}+\Delta_{\boldsymbol{k}}^{2}+t_{\boldsymbol{k}}^{2}-2t_{\boldsymbol{k}}\Delta_{\boldsymbol{k}}\cos(2k_{z}c)}\nonumber 
\end{align}
where $c$ is the interlayer spacing. $\mathcal{H}_{\boldsymbol{k}}(k_{z})$
has Weyl nodes at points $(\boldsymbol{K}_{i},0)$ where $t_{\boldsymbol{K}_{i}}=\Delta_{\boldsymbol{K}_{i}}$
and $\mathcal{E}_{\boldsymbol{K}_{i}}=\mu_{\boldsymbol{K}_{i}}$ and
a gap elsewhere provided $\left|\mu_{\boldsymbol{k}}\right|<\epsilon_{\boldsymbol{k}-}\forall\boldsymbol{k}$,
where $\epsilon_{\boldsymbol{k}\pm}=\sqrt{\mathcal{E}_{\boldsymbol{k}}^{2}+(t_{\boldsymbol{k}}\pm\Delta_{\boldsymbol{k}})^{2}}$.
If $\mu_{\boldsymbol{K}_{i}}=0$, the $i^{th}$ Weyl node is at the
Fermi level. Near $(\boldsymbol{K}_{i},0)$, $\mathcal{H}_{\boldsymbol{k}}(k_{z})-\mu_{\boldsymbol{k}}$
reduces to $H_{\text{Weyl},i}=\boldsymbol{k}_{3D}\cdot\left(\boldsymbol{v}_{i}\sigma_{z}+\boldsymbol{u}_{i}\sigma_{x}+\boldsymbol{w}_{i}\sigma_{y}\right)$,
where $\sigma_{\alpha}$ are Pauli matrices in the bilayer basis,
$\boldsymbol{v}_{i}=\boldsymbol{\nabla_{k}}\mathcal{E}_{\boldsymbol{K}_{i}}$,
$\boldsymbol{u}_{i}=\boldsymbol{\nabla_{k}}(\Delta_{\boldsymbol{K}_{i}}-t_{\boldsymbol{K}_{i}})$
and $\boldsymbol{w}_{i}=2t_{\boldsymbol{K}_{i}}c\hat{\mathbf{z}}\equiv2t_{i}c\hat{\mathbf{z}}$
are the Weyl velocities and $\boldsymbol{k}_{3D}=(\boldsymbol{k},k_{z})$.
On the top surface ($z=0$), a FA exists along the $\mathcal{E}_{\boldsymbol{k}}=\mu_{\boldsymbol{k}}$
curve where $h_{12,\boldsymbol{k}}<h_{23,\boldsymbol{k}}$ or $\Delta_{\boldsymbol{k}}<t_{\boldsymbol{k}}$.
Physically, the FA is the part of the 2D Fermi surface at $z=0$ that
survives because it has a propensity to hybridize with the missing
layer at $z=-1$.

In this model, $D_{0}=1$, so the $g_{\boldsymbol{k}}(\omega)$ is
a $c$-number that has a closed form expression in the limit $L\to\infty$
\citep{Hosur2012a}: 
\begin{align}
g_{\boldsymbol{k}}(\omega) & =\frac{1}{2t_{\boldsymbol{k}}^{2}(\omega+\mu_{\boldsymbol{k}}-\mathcal{E}_{\boldsymbol{k}})}\biggl\{(\omega+\mu_{\boldsymbol{k}})^{2}-\mathcal{E}_{\boldsymbol{k}}^{2}+t_{\boldsymbol{k}}^{2}-\Delta_{\boldsymbol{k}}^{2}\nonumber \\
 & +\sqrt{\left[(\omega+\mu_{\boldsymbol{k}})^{2}-\epsilon_{\boldsymbol{k}-}^{2}\right]\left[(\omega+\mu_{\boldsymbol{k}})^{2}-\epsilon_{\boldsymbol{k}+}^{2})\right]}\biggr\}\label{eq:Green's function}
\end{align}
The FA and LA occur at $\mathcal{E}_{\boldsymbol{k}}=\mu_{\boldsymbol{k}},t_{\boldsymbol{k}}>\Delta_{\boldsymbol{k}}$
and $\mathcal{E}_{\boldsymbol{k}}=-\mu_{\boldsymbol{k}},t_{\boldsymbol{k}}<\Delta_{\boldsymbol{k}}$,
respectively. Peeling off the $z=0$ layer corresponds to the transformation
$\mathcal{E}_{\boldsymbol{k}}\to-\mathcal{E}_{\boldsymbol{k}}$, $t_{\boldsymbol{k}}\leftrightarrow\Delta_{\boldsymbol{k}}$
in the semi-infinite limit, which interchanges the FAs and LAs as
depicted in Fig. \ref{fig:bilayer model}. Physically, the FA now
exists along the part of the $z=1$ Fermi surface that lacks a hybridization
partner. Thanks to the square root, $g_{\boldsymbol{k}}(\omega)$
clearly has branch cuts when $\omega$ is real and lies within the
bulk bands, i.e., $\epsilon_{\boldsymbol{k}-}<|\omega+\mu_{\boldsymbol{k}}|<\epsilon_{\boldsymbol{k}+}$.
In App. \ref{sec:nbc-integrals-A}, we show that the branch cut reproduces
the expected particle density on the surface due to the bulk states.

\emph{Implications for Luttinger's theorem:-} Luttinger's theorem
is a hallmark of Fermi liquid theory. It states that the Luttinger
volume -- defined as the volume enclosed by the locus of poles of
the electron Green's function equals the density of spinful (spinless)
electrons in a metal modulo 2 (modulo 1) in units of $(2\pi)^{D}$
in $D$ spatial dimensions. Importantly, the theorem dictates that
the Luttinger volume remains unchanged in the presence of interactions
\citep{Blagoev1997,Luttinger1960,Oshikawa2000,Seki2017,Voit_1995,Yamanaka1997,Heath_2020}.
It was later generalized to include Mott insulators with particle-hole
symmetry \citep{Dzyaloshinskii2003,Seki2017,Stanescu2007,Yamanaka1997,Heath_2020}.
Here, a divergent self-energy produces a Luttinger surface that encloses
a volume equal to the particle density modulo 2. In certain strongly
interacting systems, Luttinger's theorem holds in a ``soft'' form
as the Luttinger volume equals a fraction of the particle density
modulo 2 \citep{Phillips2013,Padhi:2018aa,Stanescu2007,Limtragool2018,Dave2013}.
Since there is no well-defined volume enclosed by the FAs alone in
WSMs, Luttinger's theorem is naïvely inapplicable. However, the discovery
of LAs in this work raises the question, ``does the area enclosed
by the Fermi-Luttinger loop act as a Luttinger volume and equal the
surface particle density?''

Using the minimal model described above, we prove that the answer
is negative and quantify the violation of Luttinger's theorem. Restricting
to $\mu_{\boldsymbol{k}}=0$ for simplicity, the surface particle
density at zero temperature is given by $n_{s}=-2\text{Im}\intop_{\boldsymbol{k},\omega}\Theta(-\omega)g_{\boldsymbol{k}}(\omega+i\eta)=n_{\text{s}}^{\text{p}}+n_{\text{s}}^{\text{bc}}$
where 
\begin{align}
n_{\text{s}}^{\text{p}} & =\intop_{\boldsymbol{k}}\Theta(-\mathcal{E}_{\boldsymbol{k}})R\left(1-\frac{\Delta_{\boldsymbol{k}}^{2}}{t_{\boldsymbol{k}}^{2}}\right)\label{eq:n-surf-pole}\\
n_{\text{s}}^{\text{bc}} & =\intop_{\boldsymbol{k},\omega}\Theta(-\omega)\text{sgn}(\omega)\frac{\sqrt{R\left[\left(\omega^{2}-\epsilon_{\boldsymbol{k}-}^{2}\right)\left(\epsilon_{\boldsymbol{k}+}^{2}-\omega^{2}\right)\right]}}{t_{\boldsymbol{k}}^{2}(\omega-\mathcal{E}_{\boldsymbol{k}})}\label{eq:n-surf-bc}
\end{align}
denote contributions from the poles and the branch cuts of $g_{\boldsymbol{k}}(\omega)$,
respectively, $\intop_{\boldsymbol{k},\omega}\equiv\int\frac{d^{2}kd\omega}{(2\pi)^{3}}$
and $R(x)=x\Theta(x)$. In other words, $n_{\text{s}}^{\text{p}}$
captures FA contributions to $n_{\text{s}}$ and resembles the expression
for the carrier density in a Fermi gas, but for the weight factor
$W_{\boldsymbol{k}}=R(1-\Delta_{\boldsymbol{k}}^{2}/t_{\boldsymbol{k}}^{2})$
that accounts for the varying weight of FA states on the surface and
restricts contributions to the region containing FA's, namely, $\Delta_{\boldsymbol{k}}<t_{\boldsymbol{k}}$.
In contrast, $n_{\text{s}}^{\text{bc}}$ captures bulk contributions
and approximates at $\mu=0$ to (App. \ref{sec:nbc-integrals-B}):
\begin{align}
n_{\text{s}} & =\intop_{\boldsymbol{k}}\Theta(-\mathcal{E}_{\boldsymbol{k}})+O(t,\Delta)\nu_{2D}\label{eq:weak Luttinger}
\end{align}
where $\nu_{2D}$ is the density of states at $\mathcal{E}_{\boldsymbol{k}}=0$
for the 2D layers. The first term, $\intop_{\boldsymbol{k}}\Theta(-\mathcal{E}_{\boldsymbol{k}})$,
is precisely the area enclosed by the Fermi-Luttinger loop in units
of $(2\pi)^{2}$ and can be viewed as a Luttinger volume (yellow region
in Fig. \ref{fig:k-loops-mu-0}). Physically, the equivalence at zeroth
order in the interlayer tunnelings is a remnant of Luttinger's theorem
that holds exactly for the 2D metals that form the building blocks
of the WSM. The leading violation is defined by the second term: it
is order the typical interlayer tunneling and comes from $\boldsymbol{k}$-space
regions near the Fermi-Luttinger loop, hence scaling as its circumference.

In conclusion, we have unearthed various novel features hidden in
the surface single-particle Green's function of non-interacting WSMs.
We have shown that the surface hosts LAs, normally found only in strongly
interacting systems, in addition to FAs. When the Weyl nodes are undoped,
the FAs and LAs form closed loops on the surface. Interestingly, the
LA turns into a FA when the surface layer is removed, which allows
a precise determination of the LAs. We use this principle to determine
LAs in Co$_{3}$Sn$_{2}$S$_{2}$. Finally, we showed that doping
the Weyl nodes exposes branch cuts in the Green's function that capture
the surface presence of the bulk bands. 
\begin{acknowledgments}
We thank Hridis Pal for invaluable discussions. We acknowledge financial
support from and NSF-DMR-2047193. 
\end{acknowledgments}

\begin{widetext}

\appendix
%dummy comment inserted by tex2lyx to ensure that this paragraph is not empty

\section{$n_{\text{s}}^{\text{bc}}$ integral near Weyl nodes, with doping
\label{sec:nbc-integrals-A}}

In this section, we evaluate $n_{\text{s}}^{\text{bc}}$ at non-zero
but small $\mu_{\boldsymbol{k}}$ by linearizing around each Weyl
node and compare it with the expected contribution of the bulk states
to the surface particle density. We have 
\begin{equation}
n_{\text{s}}^{\text{bc}}=\intop_{\boldsymbol{k},\omega}\Theta(\mu_{\boldsymbol{k}}-\omega)\text{sgn}(\omega)\frac{\sqrt{R\left[\left(\omega^{2}-\epsilon_{\boldsymbol{k}-}^{2}\right)\left(\epsilon_{\boldsymbol{k}+}^{2}-\omega^{2}\right)\right]}}{t_{\boldsymbol{k}}^{2}(\omega-\mathcal{E}_{\boldsymbol{k}})}
\end{equation}
where $\epsilon_{\boldsymbol{k}\pm}=\sqrt{\mathcal{E}_{\boldsymbol{k}}^{2}+(t_{\boldsymbol{k}}\pm\Delta_{\boldsymbol{k}})^{2}}$.
The integrand is non-zero only when $\epsilon_{\boldsymbol{k}-}<|\omega|<\epsilon_{\boldsymbol{k}+}$,
which defines a region around each Weyl node within an energy $|\omega|$
from the node. As long as the contributing regions around different
nodes do not overlap in $\boldsymbol{k}$-space, the $\boldsymbol{k}$-integral
can be split into integrals around each node: 
\begin{equation}
n_{\text{s}}^{\text{bc}}=\intop_{\omega}\sum_{i}\intop_{\boldsymbol{k}\approx\boldsymbol{K}_{i}}\Theta(\mu_{\boldsymbol{k}}-\omega)\text{sgn}(\omega)\frac{\sqrt{R\left[\left(\omega^{2}-\epsilon_{\boldsymbol{k}-}^{2}\right)\left(\epsilon_{\boldsymbol{k}+}^{2}-\omega^{2}\right)\right]}}{t_{\boldsymbol{k}}^{2}(\omega-\mathcal{E}_{\boldsymbol{k}})}
\end{equation}
where $\boldsymbol{K}_{i}$ are the locations of the Weyl nodes. Near
each node, let us linearize as $\mathcal{E}_{\boldsymbol{k}}\approx\boldsymbol{k}\cdot\boldsymbol{v}_{i}$,
$t_{\boldsymbol{k}}\approx t_{i}-\frac{1}{2}\boldsymbol{k}\cdot\boldsymbol{u}_{i}$,
$\Delta_{\boldsymbol{k}}\approx t_{i}+\frac{1}{2}\boldsymbol{k}\cdot\boldsymbol{u}_{i}$
and assume $2t_{i}\gg\equiv|\mu_{i}|$, where $\mu_{i}$ is the Fermi
level relative to the $i^{th}$ Weyl node. For simplicity, we ignore
tilting of the linear dispersion by neglecting linear terms in the
Taylor expansion of $\mu_{\boldsymbol{k}}$ around $\boldsymbol{K}_{i}$.
Subtracting off an infinite contribution from $\mu_{i}=0$, we get
\begin{equation}
\Delta n_{\text{s}}^{\text{bc}}=\sum_{i}\intop_{0}^{|\mu_{i}|}\frac{d\omega}{2\pi}\intop_{\sqrt{(\boldsymbol{k}\cdot\boldsymbol{v}_{i})^{2}+(\boldsymbol{k}\cdot\boldsymbol{u}_{i})^{2}}<\omega}\frac{d^{2}k}{(2\pi)^{2}}\frac{2\sqrt{\omega^{2}-(\boldsymbol{k}\cdot\boldsymbol{v}_{i})^{2}-(\boldsymbol{k}\cdot\boldsymbol{u}_{i})^{2}}}{t_{i}\left[\omega\text{sgn}(\mu_{i})-\boldsymbol{k}\cdot\boldsymbol{v}_{i}\right]}
\end{equation}
The integral simplifies upon absorbing the velocities into the momenta
as $q_{v}=\boldsymbol{k}\cdot\boldsymbol{v}_{i}=q\cos\phi$, $q_{u}=\boldsymbol{k}\cdot\boldsymbol{u}_{i}=q\sin\phi$:
\begin{align}
\Delta n_{\text{s}}^{\text{bc}} & =\sum_{i}\intop_{0}^{|\mu_{i}|}\frac{d\omega}{2\pi}\frac{2}{t_{i}u_{i}v_{i}}\intop_{q<|\omega|}\frac{qdqd\phi}{(2\pi)^{2}}\frac{\sqrt{\omega^{2}-q^{2}}}{\omega\text{sgn}(\mu_{i})-q\cos\phi}\\
 & =\sum_{i}\frac{2\text{sgn}(\mu_{i})}{t_{i}u_{i}v_{i}}\intop_{0}^{|\mu_{i}|}\frac{d\omega}{2\pi}\intop_{0}^{\omega}\frac{qdq}{2\pi}\\
 & =\frac{2}{3}\pi\sum_{i}\left(\frac{\mu_{i}}{2\pi}\right)^{3}\frac{1}{t_{i}u_{i}v_{i}}
\end{align}
In comparison, the change in the 3D bulk carrier density around a
Weyl node with velocities $u_{i}$, $v_{i}$ and $w_{i}$ due to a
local chemical potential $\mu_{i}$ follows easily from the volume
of an ellipsoid: 
\begin{equation}
\Delta N_{i}^{\text{Weyl}}=\frac{4}{3}\pi\left(\frac{\mu_{i}}{2\pi}\right)^{3}\frac{1}{u_{i}v_{i}w_{i}}\label{eq:n-Weyl-mu}
\end{equation}
Recalling that $w_{i}=2t_{i}c$, we find $\Delta n_{\text{s}}^{\text{bc}}=c\sum_{i}\Delta N_{i}^{\text{Weyl}}$,
i.e., $\Delta n_{\text{s}}^{\text{bc}}$ reflects the change in the
bulk carrier density uniformly distributed across the layers. Thus,
the branch cut in $g_{\boldsymbol{k}}(\omega)$ is a manifestation
of the continuum of bulk quasiparticle poles at fixed $\boldsymbol{k}$
and varying $k_{z}$.

\section{$n_{\text{s}}^{\text{bc}}$ integrals at general $\boldsymbol{k}$,
no doping \label{sec:nbc-integrals-B}}

In this section, we separate evaluate contributions from $\boldsymbol{k}$-points
where $\mathcal{E}_{\boldsymbol{k}}<0$ ($n_{\text{s},-}^{\text{bc}}$),
$\mathcal{E}_{\boldsymbol{k}}\approx0$ ($n_{\text{s},0}^{\text{bc}}$)
and $\mathcal{E}_{\boldsymbol{k}}>0$ ($n_{\text{s},+}^{\text{bc}}$)
to obtain $n_{\text{s}}^{\text{bc}}=n_{\text{s},-}^{\text{bc}}+n_{\text{s},0}^{\text{bc}}+n_{\text{s},+}^{\text{bc}}$
at $\mu_{\boldsymbol{k}}=0$, as needed for the quantifying the violation
of Luttinger's theorem based on the area enclosed by the Fermi-Luttinger
loop.

\subsection{$n_{\text{s},-}^{\text{bc}}$}

If $\left|t_{\boldsymbol{k}},\Delta_{\boldsymbol{k}}\right|<-\mathcal{E}_{\boldsymbol{k}}$,
then $\omega\approx\mathcal{E}_{\boldsymbol{k}}$ over the range of
the $\omega$-integral and $1/(\omega-\mathcal{E}_{\boldsymbol{k}})=2\omega/(\omega^{2}-\mathcal{E}_{\boldsymbol{k}}^{2})+O\left(\frac{t^{2},\Delta^{2},t\Delta}{\mathcal{E}^{2}}\right)$.
The leading order term is a straightforward elliptic integral in terms
of $x=\omega^{2}-\mathcal{E}_{\boldsymbol{k}}^{2}$: 
\begin{align}
n_{\text{s},-}^{\text{bc}}(0) & =\intop_{\boldsymbol{k}}\frac{\Theta(-\mathcal{E}_{\boldsymbol{k}})}{2\pi t_{\boldsymbol{k}}^{2}}\intop_{t_{\boldsymbol{k}-}^{2}}^{t_{\boldsymbol{k}+}^{2}}dx\frac{\sqrt{\left(x-\left|t_{\boldsymbol{k}}-\Delta_{\boldsymbol{k}}\right|^{2}\right)\left(\left|t_{\boldsymbol{k}}+\Delta_{\boldsymbol{k}}\right|^{2}-x\right)}}{x}\nonumber \\
 & =\intop_{\boldsymbol{k}}\frac{\Theta(-\mathcal{E}_{\boldsymbol{k}})}{4t_{\boldsymbol{k}}^{2}}\left(\left|t_{\boldsymbol{k}}+\Delta_{\boldsymbol{k}}\right|-\left|t_{\boldsymbol{k}}-\Delta_{\boldsymbol{k}}\right|\right)^{2}\nonumber \\
 & =\int_{\boldsymbol{k}}\left(1-W_{\boldsymbol{k}}\right)
\end{align}

\subsection{$n_{\text{s},0}^{\text{bc}}$}

If $\left|t_{\boldsymbol{k}},\Delta_{\boldsymbol{k}}\right|\gg\left|\mathcal{E}_{\boldsymbol{k}}\right|$,
corresponding to the region near the Fermi and Luttinger arcs, we
can restrict to a narrow region of width $O(t,\Delta)$ around the
$\mathcal{E}_{\boldsymbol{k}}=0$ contour. This gives 
\begin{align}
n_{\text{s},0}^{\text{bc}}(0) & \approx\intop_{\boldsymbol{k}}\frac{O(t,\Delta)\delta(\mathcal{E}_{\boldsymbol{k}})}{2\pi t_{\boldsymbol{k}}^{2}}\intop_{-\epsilon_{\boldsymbol{k}+}}^{-\epsilon_{\boldsymbol{k}-}}d\omega\frac{\sqrt{\left(\omega^{2}-\left|t_{\boldsymbol{k}}-\Delta_{\boldsymbol{k}}\right|^{2}\right)\left(\left|t_{\boldsymbol{k}}+\Delta_{\boldsymbol{k}}\right|^{2}-\omega^{2}\right)}}{\omega}\nonumber \\
 & =\intop_{\boldsymbol{k}}\frac{O(t,\Delta)\delta(\mathcal{E}_{\boldsymbol{k}})}{8t_{\boldsymbol{k}}^{2}}\left(\left|t_{\boldsymbol{k}}+\Delta_{\boldsymbol{k}}\right|-\left|t_{\boldsymbol{k}}-\Delta_{\boldsymbol{k}}\right|\right)^{2}\nonumber \\
 & =\frac{1}{2}\intop_{\boldsymbol{k}}O(t,\Delta)\delta(\mathcal{E}_{\boldsymbol{k}})(1-W_{\boldsymbol{k}})
\end{align}
Due to the $\delta$-function, this term receives contributions from
regions near the Fermi-Luttinger loop only. It is $O(t,\Delta)\nu_{2D}$
and constitutes the leading violation of Luttinger's theorem based
on the Fermi-Luttinger loop.

\subsection{$n_{\text{s},+}^{\text{bc}}$}

If $0<\left|t_{\boldsymbol{k}},\Delta_{\boldsymbol{k}}\right|\ll\mathcal{E}_{\boldsymbol{k}}$,
we can approximate $1/(\omega-\mathcal{E}_{\boldsymbol{k}})\approx1/2\omega$.
This gives an elliptic integral in terms of $x=\omega^{2}$ which
evaluates to 
\begin{align}
n_{\text{s},+}^{\text{bc}}(0) & \approx\intop_{\boldsymbol{k},\mathcal{E}\gg|t,\Delta|}\intop_{\epsilon_{\boldsymbol{k}-}^{2}}^{\epsilon_{\boldsymbol{k}+}^{2}}dx\frac{\sqrt{(x-\epsilon_{\boldsymbol{k}-}^{2})(\epsilon_{\boldsymbol{k}+}^{2}-x)}}{8\pi t_{\boldsymbol{k}}^{2}x}\nonumber \\
 & \approx-\intop_{\boldsymbol{k},\mathcal{E}\gg|t,\Delta|}\frac{\Delta_{\boldsymbol{k}}^{2}}{4\mathcal{E}_{\boldsymbol{k}}^{2}}
\end{align}

 \end{widetext}

\bibliographystyle{apsrev4-2}
\bibliography{library2}

\end{document}